\documentclass{article}
\usepackage[T2A]{fontenc}
\usepackage[cp1251]{inputenc}
\usepackage[english,russian]{babel}
\usepackage{amssymb,amsfonts,amsmath,mathtext}

\begin{document}

\begin{center}
{\bf\Large Configuratrix and resultant}\\[3pt]

\begin{center}
N.S. Perminov
\end{center}

\begin{center}
\textit{Kasan State University, Kazan, Russia,\\
Physical Department, Division of Relativity Theory and Gravity  }
\end{center}

\vskip12pt
\parbox{14cm}{\small In this paper, we obtain an explicit expression for the resultant of $n$ quadratic algebraic equations $\{\partial_{1}S = 0, \ldots, \partial_{n}S=0\}$, where $S$ is a cubic polynomial in $n$ variables, symmetric under permutations of its arguments. Application of this result to the study of Finslerian spaces is discussed.
}
\end{center}

%%%%%%%%%%%%%%%%%%%%
\section{Introduction} %
%%%%%%%%%%%%%%%%%%%%

In modern theoretical physics, the study of spaces with Finslerian metric functions of polynomial type \cite{Rund} becomes increasingly popular. In particular, the indicatrix equations $L(\xi)=1$ for the Berwald-Moor and Minkowski spaces can be written in polynomial form: 

\begin{align}\label{Berwald-Moor}
L^4(\xi) = \xi^{1}\xi^{2}\xi^{3}\xi^{4}=1
\end{align}
and

\begin{align}\label{Minkowski}
L^2(\xi) = \xi^{1}\xi^{2}+\xi^{1}\xi^{3}+\xi^{1}\xi^{4}+\xi^{2}\xi^{3}+\xi^{2}\xi^{4}+\xi^{3}\xi^{4}=1,
\end{align}
respectively. Of interest is also the configuratrix equation $\Phi(y)=1$, i.e, the equation $L(\xi)=1$ expressed through canonically conjugate variables $y_{i}=\partial_{i}L$. The configuratrix equation can be expressed as the condition of solvability for the following system of algebraic equations:

\begin{align}
\left\{
\begin{array}{lll}
L(\xi)=1,\\
\\
\dfrac{\partial}{\partial\xi^{1}}L(\xi)=y_{1} , \\
\\
...\\
\\
\dfrac{\partial}{\partial\xi^{n}}L(\xi)=y_{n}.
\end{array}
\right.
\end{align}
It is well known, that a system of polynomial equations is consistent if and only if certain expression, called \emph{resultant} \cite{Gelf, MD, MSh} vanishes. Accordingly, when $L^{k}(\xi) \equiv S(\xi)$ is a polynomial of $\xi$, the configuratrix equation is a condition of vanishing resultant:

\begin{align}
R \left \{
\begin{array}{c}
S(\xi)-1\\
\\
\dfrac{\partial}{\partial\xi^{1}}S(\xi)-k y_{1}\\
\\
...\\
\\
\dfrac{\partial}{\partial\xi^{n}}S(\xi)-k y_{n}
\end{array}
\right \} = 0.
\end{align}
This connection between the configuratrix equation and resultant theory opens new possibilities: in some cases, it is possible to evaluate resultants without solving directly the non-linear algebraic equations. 

Another topic in geometry of Finslerian spaces, which is also closely related to resultants, is the possibility to introduce a renormalisable volume element. It is related to finiteness of the region, bounded by the indicatrix $L(\xi)=1$. Whether this region is finite or not, depends on existence of critical points on the surface $L(\xi)=0$, i.e, depends on solvability of the system $\{\partial_{i}L=0\}$. This system is solvable, if and only if

\begin{align}\label{Resultant}
R\{\partial_{i}S\}=0,
\end{align}
where $L^k (\xi)=S(\xi)$ is a polynomial. As we can see, resultant can have many applications in Finslerian geometry. For this reason, it is important to find simple and explicit formulas for resultants in various particular cases. In this paper, we consider the case when $S(\xi)$ is a cubic polynomial (polynomial of degree $3$) of $n$ variables, symmetric under permutations of all its arguments.

%%%%%%%%%%%%%%%%%%%%%%%%%%%%%%%%%%
\section{Calculation of the resultant} %
%%%%%%%%%%%%%%%%%%%%%%%%%%%%%%%%%%

Any symmetric polynomial of degree 3 can be uniquely represented as

\begin{align}\label{SymPol[3,n]}
S(x_{1},...,x_{n}) =A_{1}(s_{1})^{3}+A_{2}s_{1}s_{2}+A_{3}s_{3},
\end{align}
where $\{s_{1},...,s_{n}\}$ are the elementary symmetric polynomials,

$$
s_{1}=\sum\limits_{i} x_i,
$$

$$
s_{2}=\sum\limits_{i<j} x_i x_j
$$

$$
\ldots
$$

\begin{align}\label{elemsympol}
s_{k}=\sum\limits_{i_{1}<...<i_{k}} x_{i_{1}}...x_{i_{k}}.
\end{align}
In this case, we have

\begin{align}\label{F_{i}}
\dfrac{\partial}{\partial x_{i}}S=A_{3}x_{i}^{2}-(A_{2}+A_{3})x_{i}s_{1}+(3A_{1}+A_{2})s_{1}^{2}+(A_{2}+A_{3})s_{2}.
\end{align}
After a linear transformation

\begin{align}\label{FF}
F_{i}=\dfrac{1}{A_{3}}\partial_{i}S+\dfrac{A_{2}+A_{3}}{A_{3}(2A_{3}-n(A_{2}+A_{3}))}\sum\limits_{i = 1}^{n} \partial_{i}S
\end{align}
we obtain another system of equations, equivalent to (\ref{F_{i}}):

\begin{align}\label{FFF}
F_{i}=x_{i}^{2}+2Ax_{i}s_{1}+Bs_{1}^2,
\end{align}
where

\begin{align}
\left\{
\begin{array}{lll}
A=-\dfrac{A_{2}+A_{3}}{2A_{3}},\\
\\
B=\dfrac{6A_{1}A_{3}+2A_{2}A_{3}-A_{2}^{2}}{A_{3}(2A_{3}-n(A_{2}+A_{3}))} .
\end{array}
\right.
\end{align}
The resultants of these two systems are related \cite{MD}:

\begin{align}\label{R1=f(R2)}
R\{\partial_{i}S\}=R\{F_{i}\}\left(\dfrac{A_{3}^{n-1}(2A_{3}-n(A_{2}+A_{3}))}{2}\right)^{2^{n-1}} .
\end{align}
Using the Poisson product formula \cite{Gelf, MSh} we obtain

\begin{align}
\begin{array}{ccc}
\label{solvation} R\big(x_{i}^{2}+2Ax_{i}s_{1}+Bs_{1}^2\big)=\\
\\
=\prod\limits_{k_{1},...,k_{n} = 0}^{1} \det_{n \times n}\big( \delta_{ij} + \lambda (-1)^{k_i} \delta_{ij} + A E_{ij} \big)= \\
\\
= \prod\limits_{k_{1},...,k_{n} = 0}^{1} \big(1 + nA + \lambda \sum\limits_{j = 1}^{n} (-1)^{k_{j}} \big),
\end{array}
\end{align}
where

$$ \delta_{ij} = \left\{ \begin{array}{cc} 1, \ \mbox{если } i = j \\ 0, \ \mbox{ если } i \neq j \end{array} \right., \ \ E_{ij} = 1, \ \ \ \lambda^2 = A^2 - B. $$
It is easy to show, that this expression can be simplified to a form

\begin{align}\label{R final 1}
\prod\limits_{k = 0}^{n-1} \Big[ (1+nA)^2+(A^{2}-B)(n-2k)^2\Big]^{C_{n-1}^{k}},
\end{align}
where

$$C_{p}^{q}=\dfrac{p!}{q!(p-q)!}$$
are the binomial coefficients. After simple algebraic calculations, we obtain 

\begin{align}\label{R final 2}
R\{\partial_{i}S\}=\prod\limits_{k = 0}^{n-1}Y_{k}^{C_{n-1}^{k}},
\end{align}

\begin{align}\label{R final 3}
\nonumber Y_{k}=\dfrac{1}{4}A_{3}^{n-3}\Big\{ (2A_{3}-n(A_{2}+A_{3}))^{3}-(n-2k)^{2} \cdot \\
\cdot ( (A_{2}+A_{3})^{2}(2A_{3}-n(A_{2}+A_{3}))-4A_{3}(6A_{1}A_{3}+A_{2}A_{3}-A_{2}^{2}) )\Big\}
\end{align}
This result can be recast in even simpler form, by doing a linear change of parameters

\begin{align}
\left\{
\begin{array}{lll}
B_{1}=n^{2}A_{1}+\dfrac{n(n-1)}{2}A_{2}+\dfrac{(n-1)(n-2)}{6}A_{3},\\
\\
B_{2}=nA_{2}+(n-2)A_{3},\\
\\
B_{3}=A_{3},
\end{array}
\right.
\end{align}
We finally obtain

\begin{equation}
\addtolength{\fboxsep}{5pt}
\boxed{
\begin{gathered}
R\big\{ \partial_i S \big\} = \big( B_3 \big)^{(n-3)2^{n-1}} \prod\limits_{k = 0}^{n - 1} \left( \dfrac{6(n-2k)^2}{n^2} B_1 B_3^2 - \dfrac{k(n-k)}{n^2} B_2^3 \right)^{C_{n-1}^{k}}
\end{gathered}
}\label{mainresult}
\end{equation}
This simple and explicit formula is the main result of present paper.

%%%%%%%%%%%%%%%%%%%%%%%%%%%%%%%%%%
\section{Conclusion} %
%%%%%%%%%%%%%%%%%%%%%%%%%%%%%%%%%%

The use of resultants \cite{Gelf, MD, MSh} in Finslerian geometry can simplify the understanding of some classes of metric structures. The expression (\ref{mainresult}), which we obtained, gives a simple and efficient way to calculate the resultant of a system $\{\partial_{i}S=0\}$
for polynomial $S$ of degree 3 in any dimension. In the future we will try to generalise (\ref{mainresult}) and obtain analogous formulas for symmetric polynomials $S$ of higher degrees $r > 3$.


\begin{thebibliography}{par}
%--------------------------------------------------------------------
\bibitem{Rund} H. Rund, {\em Differential geometry of Finslerian spaces}, М.: Nauka, 1981.
%--------------------------------------------------------------------
\bibitem{Gelf} M. Gelfand, M.M. Kapranov, and A.V. Zelevinsky, {\em
Discriminants, Resultants and Multidimensional Determinants},
Birkhauser, 1994.
%--------------------------------------------------------------------
\bibitem{MD} V.Dolotin and A.Morozov, \emph{Introduction to Non-Linear Algebra}, World Scientific, 2007, hep-th/0609022
%--------------------------------------------------------------------
\bibitem{MSh}A.Morozov and Sh.Shakirov, \emph{Analogue of the identity Log Det = Trace Log for resultants}, arXiv:0804.4632 \\
%--------------------------------------------------------------------
\end{thebibliography}
\end{document}